\documentclass[12pt,english,floatfix,showkeys,superscriptaddress,aps,prd,preprint]{revtex4}
\usepackage[latin1]{inputenc}
\usepackage[T1]{fontenc}
\usepackage{lmodern}
\usepackage{amsmath}
\usepackage{amssymb}
\usepackage{graphicx}
\usepackage{float}
\usepackage{esint}
\usepackage{longtable}
\usepackage{dcolumn}
\usepackage{babel}
\usepackage{csquotes}
\usepackage{color}
\usepackage{slashed}
\usepackage{simplewick}
\usepackage{amsmath,latexsym}

\usepackage{hyperref}
\hypersetup{
    colorlinks,
    citecolor=blue,
    filecolor=green,
    linkcolor=purple,
    urlcolor=red,
}

\usepackage{slashed}

\usepackage{hyperref}
\hypersetup{colorlinks,breaklinks,
			citecolor=[rgb]{0,0.0,1.0},
            urlcolor=[rgb]{0.0,0.0,1.0},
            linkcolor=[rgb]{0,0.5,0.9}}

\begin{document}

\title{Regular Black Holes in $D = 2+1$ with $f(R)$ Gravity}

\author{A. C. L. Santos}
\email{alanasantos@fisica.ufc.br}
\affiliation{Universidade Federal do Cear\'a (UFC), Departamento de F\'isica,\\ Campus do Pici, Fortaleza - CE, C.P. 6030, 60455-760 - Brazil.}
\author{R. V. Maluf}
\email{r.v.maluf@fisica.ufc.br}
\affiliation{Universidade Federal do Cear\'a (UFC), Departamento de F\'isica,\\ Campus do Pici, Fortaleza - CE, C.P. 6030, 60455-760 - Brazil.}
\affiliation{Departamento de F\'{i}sica Te\'{o}rica and IFIC, Centro Mixto Universidad de Valencia - CSIC. Universidad
de Valencia, Burjassot-46100, Valencia, Spain.}

\author{C. R. Muniz}
\email{celio.muniz@uece.br}
\affiliation{Universidade Estadual do Cear\'a (UECE), Faculdade de Educa\c{c}\~ao, Ci\^encias e Letras de Iguatu, Av. D\'ario Rabelo s/n, Iguatu-CE, 63.500-00 - Brazil.}


\date{\today}

\begin{abstract}
Despite the experimental success of general relativity in the scales it has been tested, there are still some inconsistencies, such as explaining the acceleration of the universe and singularities. $f(R)$ gravity has emerged to provide corrections to the Ricci scalar, leading to new equations of motion that could explain this acceleration. On the other hand, Regular Black Holes, supported by non-linear electrodynamics, have well-defined curvature scalars but often violate energy conditions. In this work, we investigate how to obtain regular solutions in $(2+1)$ dimensions by considering $f(R)$ gravity and non-linear electrodynamics. Subsequently, we find that non-linear electrodynamics models that guarantee regular solutions in General Relativity do not produce regular solutions in $f(R)$ theory. Finally, we observe that energy conditions are still violated in $f(R)$ gravity.
\end{abstract}


\maketitle


\section{Introduction}
The experimental success of General Relativity at the scales it has been tested is undeniable. From classical tests (such as the accurate prediction of light deflection and the explanation for the precession of Mercury, for instance \cite{Will:2005va}), to the prediction of Black Holes and Gravitational Waves, which were recently confirmed \cite{EventHorizonTelescope:2019dse, EventHorizonTelescope:2022wkp, LIGOScientific:2016aoc}, the theory presents considerable robustness. However, some aspects remain misunderstood: the confirmed accelerated expansion of the universe \cite{SupernovaSearchTeam:1998fmf} and singularities \cite{Penrose:1964wq}. Over the years, various theoretical approaches have emerged in an attempt to shed light on these questions.

In the search for corrections to the equations of motion to explain the accelerated expansion of the universe, extensions or modifications of General Relativity have emerged -- since the beginning of the theory -- such as the inclusion of Torsion and non-metricity (a great compilation can be found in \cite{CANTATA:2021ktz}). In this way, $f(R)$ gravity \cite{Olmo:2011uz} predicts corrections given by nonlinear terms of the curvature scalar in the Lagrangian that would lead to cosmological self-accelerating solutions without resorting to the existence of dark energy. With this framework, this extension has been extensively studied in astrophysical and cosmological scenarios \cite{delaCruz-Dombriz:2006kob, Li:2007xn, Bamba:2013fha, Capozziello:2008ch, Muniz:2022eex}. Also, an interesting feature of this model is the mapping into a scalar-tensor theory, which has also undergone deep investigation in cosmological scenarios \cite{Olmo:2005hc}.

On the other hand, since Bardeen's pioneering solution \cite{BARDEEN}, Regular Black Holes have emerged in a classical context: Black Holes that possess well-defined curvature scalars at every point in space \cite{Lan:2023cvz}. These solutions are obtained when we consider nonlinear electrodynamics as source \cite{Ayon-Beato:2000mjt} or fluids with certain specific energy densities \cite{Maluf:2022jjc}. The cost of this regularity is the violation of energy conditions, most of the time. In this sense, many regular solutions have been investigated in the literature \cite{Rodrigues:2020pem, Olmo:2023lil, Guerrero:2022msp}, including quantum corrections \cite{Maluf:2018ksj}.

Despite some problems, like the absence of dynamical degrees of freedom, since the first Black Hole solution in ($2+1$) dimensions found by Ba\~{n}ados, Teitelboim and Zanelli \cite{Banados:1992wn}, the interest in solutions in low-dimensional contexts has been expanding \cite{Karakasis:2021lnq, Karakasis:2021ttn}. Among the reasons, we can emphasize the various predicted and detected effects in lower dimensions in other branches, such as Statistical Mechanics and Quantum Mechanics. It is worth mentioning that the unexpected and intriguing Unruh effect was initially predicted in $(1+1)$ dimensions \cite{Unruh:1976db} and the most promising experimental investigation of the EP=EPR conjecture was carried out in a two-dimensional gravity dual system \cite{Jafferis:2022crx}. Also, the possibility of spontaneous dimensional reduction is predicted in some models to occur at extremely high energy scales \cite{Carlip:2016qrb}.

In this context, to address a solution that deals with both the accelerated expansion of the universe and singularities, we dare to ask: Just as in ($3+1$) dimensions \cite{Rodrigues:2015ayd, Rodrigues:2016fym}, do we find regular solutions in ($2+1$) dimensions in $f(R)$ gravity? If so, what is the most general method to ensure that we will have a regular solution? Do nonlinear electrodynamics and energy densities that ensured regular solutions in General Relativity still ensure regular solutions in $f(R)$ gravity? Are the energy conditions still violated?

To answer these questions, we have organized the paper as follows: in section \ref{section II}, without specifying a $f(R)$ model, we find the necessary conditions for having a regular solution supported by non-linear electrodynamics. Subsequently, we consider models of non-linear electrodynamics that guarantee regular solutions in General Relativity to investigate if these models still support regular solutions in a $f(R)$ gravity and we also study the energy conditions. In section \ref{section III}, considering a regular geometry, we investigate whether the energy conditions are violated. Finally, in section \ref{conclusion}, we outline our perspectives and conclusions.

\section{General Regular Black Holes Solutions With NED}\label{section II}
In this section, we are interested in creating the most general way to find a regular solution in a $f(R)$ theory in ($2+1$) dimensions considering nonlinear electrodynamics. In this sense, we start with the following action
\begin{equation}\label{action}
S=\int d^3x\sqrt{-g}\left[\frac{f(R)}{16\pi} + L(F)\right],    
\end{equation}
where $g$ stands for the determinant of the metric $g_{\mu\nu}$, $f(R)$ is an arbitrary function of the Ricci scalar and $L(F)$ is the Lagrangian of the nonlinear electrodynamics with $F = F^{\mu\nu}F_{\mu\nu}$. Varying (\ref{action}) with respect to the metric, we obtain the equations of motion
\begin{equation}\label{eom}
f_R R_{\mu\nu} - \frac{1}{2}g_{\mu\nu}f(R)+g_{\mu\nu}\Box f_R - \nabla_\mu \nabla_\nu f_R = 8\pi T_{\mu\nu}, 
\end{equation}
where $f_R$ represents the derivative of $f(R)$ with respect to $R$. The energy-momentum tensor is given by
\begin{equation}\label{Tmn}
T_{\mu\nu} \equiv -\frac{2}{\sqrt{-g}}\frac{\delta(\sqrt{-g} L_m)}{\delta g^{\mu\nu}} = g_{\mu\nu}L(F)-4L_{,F}F_{\mu\alpha}F_{\nu}^{~\alpha},    
\end{equation}
and $L_{,F}$ represents the derivative of $L(F)$ with respect to $F$. The electromagnetic field equations are determined by varying with respect to the potential $A_{\mu}$, remembering that the field-strength tensor $F_{\mu\nu}$ is defined as $F_{\mu\nu}= \partial_{\mu}A_{\nu}-\partial_{\nu}A_{\mu}$, we obtain
\begin{equation}\label{em}
\nabla_\mu(L_{,F}F^{\mu\nu})=0.   
\end{equation}\label{EMeq}
We will assume a circularly symmetric spacetime
\begin{equation}\label{metric}
ds^2=-b(r)dt^2+b(r)^{-1}dr^2+r^2d\phi^2,   
\end{equation}
where $b(r)$ is an arbitrary function of the radial coordinate $r$. It is worth mentioning that the choice $ \displaystyle g_{tt}g_{rr}= -1$, it is naturally obtained in General Relativity, but that is not the case in $f(R)$ gravity theories, where such a choice represents a further constraint on the system \cite{Rodrigues:2015ayd}. Considering radial electric and scalar magnetic fields \cite{Cataldo:2000ns}, then
\begin{equation}\label{metrica3}
F_{\mu\nu} = \left(\begin{array}{cccc} 
0 & E(r) & 0 \\
-E(r) & 0 & -B(r) \\
0 & B(r) & 0
\end{array}\right).    
\end{equation}    
Using the ansatz to the metric and the energy-momentum tensor, we find
\begin{eqnarray}\label{1eq}
rf(R)+16 \pi r L(F) + b'(r) f_R(r) -2 b(r) f_R'(r)-r b'(r) f_R'(r)  \nonumber \\ + 64 \pi  r E(r)^2 L_{,F} + r f_R(r) b''(r) - 2rb(r) f_R''(r) = 0.
\end{eqnarray}
\begin{eqnarray}\label{2eq}
16 \pi r^2 L(F)+r f_R(r)b'(r)-2rb(r)f_R'(r)-r^2b'(r)f_R'(r) + r^2 f(R) \nonumber \\ + 64\pi r^2 E(r)^2L_{,F} -64\pi b(r) B(r)^2L_{,F} + r^2f_R(r)b''(r) = 0.   
\end{eqnarray}
\begin{eqnarray}\label{3eq}
- r^2 f(R) +2(-8\pi r^2L(F) - rf_R(r)b'(r)+r^2b'(r)f_R'(r) \nonumber \\  + 32\pi b(r)B(r)^2 L_{,F}+r^2b(r)f_R''(r)) = 0.   
\end{eqnarray}
\begin{equation}\label{B}
32\pi b(r) E(r) B(r) L_{,F} = 0.    
\end{equation}
Where the prime $(')$ stands for the total derivative with respect to the radial coordinate $r$. From equation (\ref{B}), we obtain that, curiously, the radial magnetic field or the radial electric field should be zero. We will consider the radial electric field just like in General Relativity \cite{Cataldo:2000ns}. Therefore, the equation of motion from the matter sector (\ref{em}) is
\begin{equation}\label{4eq}
E(r)L_{,F} = - \frac{q}{r}.    
\end{equation}
Where $q$ is an integration constant related to the electric charge. Since $B(r)=0$, subtracting (\ref{1eq}) from (\ref{2eq}), we have:
\begin{equation}\label{fr}
rb(r)f''_R(r) = 0.
\end{equation}
The solution of equation (\ref{fr}) is given by
\begin{equation}\label{fR}
f_{R} (r) = \alpha r+c_1,   
\end{equation}
where $c_1$ and $\alpha$ are constants of integration. The integration of (\ref{fR}) yields
\begin{equation}\label{f(R)}
f(R) = R-2\Lambda + \alpha \int r(R) dR,  
\end{equation}where we have set an integration constant to $-2\Lambda$ to match with General Relativity in the particular case with $\alpha=0$ \cite{Maluf:2022jjc}. And, therefore, we have corrections to Ricci scalar and consequently to General Relativity, as expected. At this point, it is interesting to note that the cosmological constant emerges naturally from the equations of motion as an integration constant of the $f(R)$ theory. On the other hand, with the metric ansatz (\ref{metric}), the invariant scalars, Ricci and Kretschmann, are given by, respectively
\begin{equation}\label{ricciscalar}
R = - b''(r)-\frac{2b'(r)}{r},    
\end{equation}
\begin{equation}
K = b''(r)^2 + \frac{2b'(r)^2}{r^2}.    
\end{equation}
To find regular solutions, we will adopt the He and Ma procedure \cite{He:2017ujy}. We begin by imposing a regular geometry, which means these scalars are well-defined in all domains, and so we should have the limit satisfied 
\begin{equation} 
\lim_{r\to 0} \frac{b'(r)}{r} = constant.
\end{equation}
Since in a vacuum solution, we need to recover the Ba\~{n}ados, Teitelboim and Zanelli case \cite{Banados:1992wn}, we take the form to the metric 
\begin{equation}\label{b(r)}
b(r) = -m + \frac{r^2}{l^2} + k(r),
\end{equation}
where $k(r)$ is a function responsible for guaranteeing the regularity of the solution and should be expanded in the following way \cite{He:2017ujy}   
\begin{equation}\label{k(r)}
k(r) = k_0 + k_2 r^2 + O(r^3),     
\end{equation}
avoiding the existence of poles. Deriving (\ref{3eq}) with respect to $r$ and replacing  (\ref{fR}), (\ref{ricciscalar}) and (\ref{b(r)}), we obtain
\begin{eqnarray}\label{condI}
-k^{(3)}(r)(1+\alpha r) - 2\alpha k''(r) + \frac{2\alpha k'(r)}{r} + 16 \pi \frac{\partial L(F)}{\partial F}\frac{\partial F}{\partial r} = 0.
\end{eqnarray}
At this point, we want to obtain more physical results imposing that the nonlinear electrodynamics should reduce to the Maxwell theory in the weak field limit, namely $L(F) \to - F$, for large $r$ \cite{Maluf:2022jjc}. So, in the asymptotic limit
\begin{equation}\label{condII}
-k^{(3)}(r)(1+\alpha r) - 2\alpha k''(r) + \frac{2\alpha k'(r)}{r} - \frac{64\pi q^2}{r^3} = 0.
\end{equation}
And when $\displaystyle \alpha\to 0$, $\displaystyle k^{(3)}(r) = \frac{64\pi q^2}{r^3}$, which it is in agreement with General Relativity, where $\displaystyle k'(r) \propto \frac{1}{r}$ and we have a cosmological constant in the action (\ref{action}) \cite{He:2017ujy}. In this way, we can construct regular black holes with functions $k(r)$ satisfying the equations (\ref{k(r)}), (\ref{condI}) and (\ref{condII}). These equations were obtained without specifying the $f(R)$ gravity. However, we naturally obtained that the constraint of the equation (\ref{fR}) should be respected. To find such a theory,  we must find the $r(R)$ based on the Ricci scalar and, finally, solve the integral (\ref{f(R)}).

\section{New Solutions Considering NED's Models}\label{section III}

First of all, let us analyze the BTZ-like charged solution, which means the case when $L(F)= - F$, where $ \displaystyle F(r)=-\frac{2q^2}{r^2}$, so we obtain:
\begin{align}\label{BTZ-like}
k(r) = 32\pi q^2\left(\ln (r) \left(\alpha ^2 r^2 \ln (r)+2 \alpha  r (\alpha  r+1)-1\right)+4 \alpha  r\right. \nonumber \\
\left.+ 2 \alpha ^2 r^2 \text{Li}_2(r \alpha +1)-2 \alpha ^2 r^2 (-\ln (\alpha  (-r))+\ln (r)+1) \ln (\alpha  r+1)\right).
\end{align}
And when $\alpha \to 0$, we obtain
\begin{equation}
k(r) = -32 \pi  q^2 \ln (r).
\end{equation}
Is it a regular solution? The Ricci scalar is given by:
\begin{eqnarray}
R &=& 192 \pi  \alpha^2 q^2 (2 (-\ln (-\alpha  r+\ln (r+1) \ln (\alpha  r+1)-\ln (r (\ln (r+2))-\frac{6}{l^2} \nonumber \\ &+& 
\frac{1}{r^2 (\alpha  r+1)^2}(32 \pi  q^2 (-12 \alpha ^2 r^2 ((\alpha  r+1)^2 \text{Li}_2(\alpha  r+1) +\alpha  r (\ln (r+2))
\nonumber \\ &-& 3 \alpha  r (13 \alpha  r+4)-2 \alpha  r (9 \alpha  r+2) \ln (r+1)).
\end{eqnarray}
In the asymptotic limit, $r \to 0$
\begin{equation}
R \approx \frac{32 \pi  q^2}{r^2},    
\end{equation}
which diverges independently of $\alpha$ and, consequently, $f(R)$ theory. Therefore, the $f(R)$ gravity does not "cure" the singularity of the solution. On the other hand, He and Ma \cite{He:2017ujy} construct in the context of General Relativity some $(2 + 1)$-dimensional regular black holes with specifics $k(r)$ functions and the associated nonlinear electrodynamics which guarantees it. As examples of the generality of the equation (\ref{condI}), we are interested in investigating if these nonlinear electrodynamics continue to ensure regular solutions in a $f(R)$ theory.

Case I:
\begin{equation}
L(r) = - \frac{a^2q^2}{16 \pi (a^2+r^2)^{\frac{3}{2}}}.    
\end{equation}
So, the equation (\ref{condI}) becomes
\begin{eqnarray}
-k^{(3)}(r)(1+\alpha r) - 2\alpha k''(r) + \frac{2\alpha k'(r)}{r} + \frac{3 a^2 q^2 r}{\left(a^2+r^2\right)^{5/2}} = 0.
\end{eqnarray}
Wich solution it is
\begin{align}\label{CaseI}
k(r) =  q^2\left(\sqrt{a^2+r^2} (2 \alpha  r-1)+r^2\alpha \left(2 a \alpha  \ln (r)-2 a \alpha  \ln \left(a \left(\sqrt{a^2+r^2}+a\right)\right)\right.\right. \nonumber \\
\left.\left.+\frac{\left(2 a^2 \alpha ^2+1\right) \left(\ln \left(\alpha  a^2+\sqrt{\left(a^2 \alpha ^2+1\right) \left(a^2+r^2\right)}-r\right)-\ln (\alpha  r+1)\right)}{\sqrt{a^2 \alpha ^2+1}}\right)\right),
\end{align}
and when $ \alpha \to 0$, we recover the General Relativity result, where:
\begin{equation}
k(r) = - q^2 \sqrt{a^2+r^2}.    
\end{equation}
The Ricci scalar is a long equation, which in the asymptotic limit $r \to 0$, we obtain the dominant term
\begin{equation}
R \approx -\frac{4 \sqrt{a^2} \alpha  q^2}{r},    
\end{equation}
that diverges. Evidently, when $\alpha \to 0$ we recover the regularity obtained in the General Relativity result.

Case II:
\begin{equation}
L(r) = \frac{a^2q^2(ar-q)}{8\pi(ar+q)^3}.    
\end{equation}
So, the equation (\ref{condI}) becomes
\begin{eqnarray}
-k^{(3)}(r)(c_1+\alpha r) - 2\alpha k''(r) + \frac{2\alpha k'(r)}{r} -\frac{4a^3 q^2 (a r-2 q)}{  (a r+q)^4} = 0.
\end{eqnarray}
Wich solution is given by
\begin{align}\label{CaseII}
k(r) = \frac{2}{3}q^2\left(\frac{3 a \left(a^2 r^2+a q r+q^2\right)}{q (\alpha  q-a) (a r+q)}-\frac{12 \alpha  q}{a}+\frac{3 a r}{q}+3 \alpha ^2 r^2 (2 \ln (q)+5) \ln (r)+15 \alpha  r+\pi ^2\right. \nonumber \\
-\frac{3 \alpha ^2 r^2 \left(2 a^2-8 \alpha  a q+5 \alpha ^2 q^2\right) \ln (\alpha  r+1)}{(a-\alpha  q)^2}-\frac{6 \ln \left(\frac{\alpha  (a r+q)}{\alpha  q-a}\right) \left(\left(a-a \alpha ^2 r^2\right) \ln (\alpha  r+1)+\alpha  q\right)}{a} \nonumber \\ 
+3\ln (a r+q)\left(-\frac{a^4 r^2}{q^2 (a-\alpha  q)^2}+\frac{a^2 r^2}{q^2}+2 \alpha  \left(\frac{a r^2}{q}+\frac{q}{a}+r\right)-2 \left(\alpha ^2 r^2-1\right) \ln (\alpha  r+1)-1\right. \nonumber \\ 
\left.\left.-2 \ln \left(\frac{\alpha  a r+a}{a-\alpha  q}\right)\right)-6 \alpha ^2 r^2 \text{Li}_2\left(-\frac{a r}{q}\right)-6 \text{Li}_2\left(\frac{(q+a r) \alpha }{q \alpha -a}\right)+6 \left(\alpha ^2 r^2-1\right) \text{Li}_2\left(\frac{r \alpha  a+a}{a-q \alpha }\right)\right).
\end{align}
And when $\alpha \to 0$, we recover the General Relativity result, where
\begin{equation}
k(r) = -\frac{2 q^3}{a r+q}-2 q^2 \ln (a r+q). 
\end{equation}
In the asymptotic limit, $r \to 0$, we have that the Ricci scalar is given by
\begin{equation}
R \approx -\frac{4 \alpha  q^2 (2 \ln (q)+5)}{r},    
\end{equation}
that diverges and also when $\alpha \to 0$ we recover the regularity obtained in the General Relativity case. In fact, the result that these solutions are not regular is not surprising, since the $k(r)$ functions obtained from the nonlinear electrodynamics do not satisfy the condition (\ref{k(r)}). Therefore, the modifications induced by a $f(R)$ theory remove the regularity of solutions conferred by this nonlinear electrodynamics in $(2+1)$ dimensions.

\subsection{Energy Conditions}
To describe the energy conditions with the appropriate physical interpretation, we follow the approach of \cite{Rodrigues:2015ayd} and rewrite the equations of motion (\ref{eom}) as
\begin{equation}
R_{\mu\nu} - \frac{1}{2}g_{\mu\nu} R = f_R^{-1}\left[ 8\pi T_{\mu\nu} + \frac{1}{2}g_{\mu\nu} (f-Rf_R) + (\nabla_{\mu} \nabla_{\nu} - g_{\mu\nu} \Box) f_R \right] \equiv 8\pi \mathcal{T}_{\mu\nu}^{(eff)},      
\end{equation}
where,
\begin{equation}
{\mathcal{T}}^{0(eff)}_0 = -\rho^{(eff)},\; {\mathcal{T}}^{1(eff)}_1 = p_{r}^{(eff)},\; {\mathcal{T}}^{2(eff)}_2 = p_{\theta}^{(eff)}.    
\end{equation}

This redefinition could be interpreted as the possibility of transferring nonlinearity from the geometric sector to the matter sector \cite{Afonso:2018hyj} to study the energy conditions in the background of the geometry of General Relativity where the theorems of singularities were described. Using the ansatz to the metric and the energy-momentum tensor, we find
\begin{eqnarray}
\rho^{(eff)} &=& \frac{-1}{16\pi rf_R(r)}(rf(R)+16\pi r L(F) +2f_R(r)b'(r)-2b(r)f_R'(r) \nonumber \\ &-& rb'(r)f_R'(r)+64\pi r E^2(r)L_F+rf_R(r)b''(r)),   
\end{eqnarray}
\begin{eqnarray}
p_{r}^{(eff)} &=&  \frac{1}{16\pi rf_R(r)}(rf(R)+16\pi r L(F) +2f_R(r)b'(r)-2b(r)f_R'(r) \nonumber \\ &-& rb'(r)f_R'(r)+64\pi r E^2(r)L_F+rf_R(r)b''(r)),    
\end{eqnarray}
\begin{eqnarray}
p_{\theta}^{(eff)} = -\frac{(rf(R)+16 \pi r L(F) + 2f_R(r)b'(r)-2rb'(r)f_R'(r)+rf_R(r)b''(r)}{16\pi rf_R(r)}.    
\end{eqnarray}
Considering (\ref{1eq}), (\ref{2eq}) and (\ref{3eq}), we can reduce to:
\begin{equation}
\rho^{(eff)} = - p_{r}^{(eff)} = - \frac{b'(r)}{16 \pi r},   
\end{equation}
\begin{equation}
p_{\theta}^{(eff)} = \frac{b''(r)}{16 \pi}.    
\end{equation}

\begin{figure}[!h]
\begin{center}
\begin{tabular}{ccc}
\includegraphics[height=4.9cm]{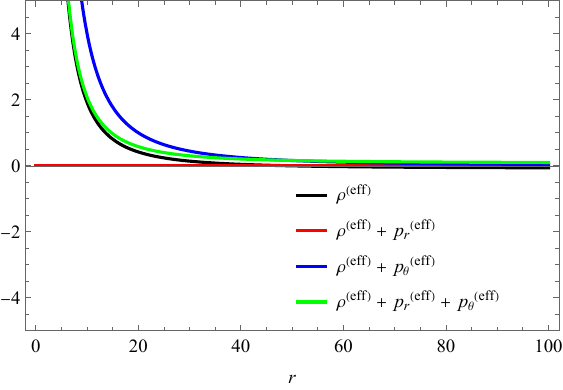}
\includegraphics[height=5cm]{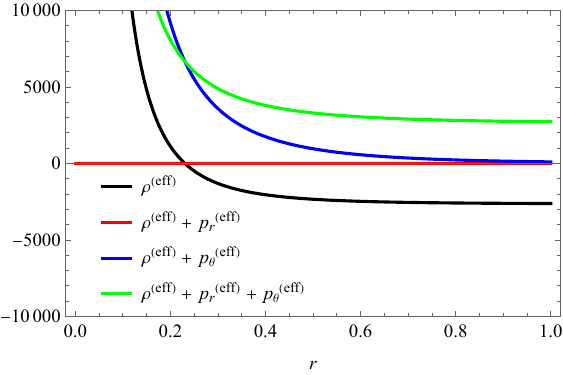}\\ 
(a) \hspace{6 cm}(b)\\
\includegraphics[height=5cm]{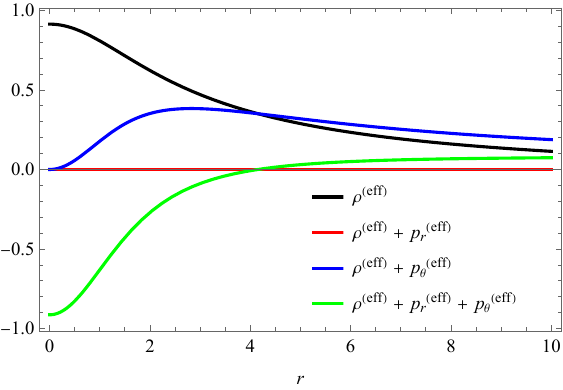}
\includegraphics[height=5cm]{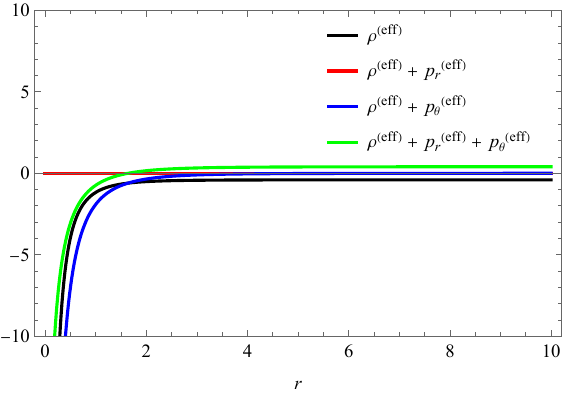}\\ 
(c) \hspace{6 cm}(d)\\
\includegraphics[height=4.9cm]{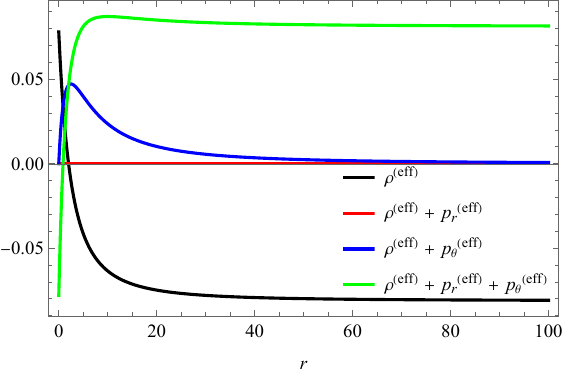}
\includegraphics[height=5cm]{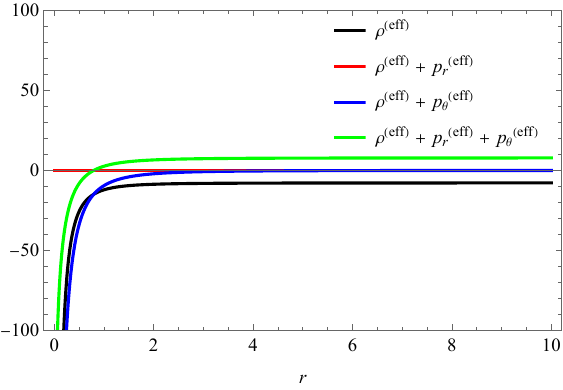}\\ 
(e) \hspace{6 cm}(f)\\
\end{tabular}
\end{center}
\caption{
The graphical representation of the radial dependence for $\rho^{(eff)}$ and combinations with $p_{r}^{(eff)}$ and $p_{\theta}^{(eff)}$ given from solutions (\ref{BTZ-like}), (\ref{CaseI}), (\ref{CaseII}), respectively, with $l=0.7, q=10, m=80, a=2, \alpha = 0$ (left panels) and $l=0.7, q=10, m=80, a=2, \alpha = 1$ (right panels).\label{Figure1}}
\end{figure}

Compatible with the relationships imposed between $\mathcal{T}_{\mu\nu}^{(eff)}$ and the Einstein tensor, $G_{\mu\nu}$. Now we can draw a parallel with General Relativity to study the energy conditions that impose: the energy density plus any of the pressures cannot be negative (Null Energy Condition), the energy density must be greater than or equal to the absolute value of any of the pressures (Dominant Energy Condition), the sum of energy density and pressure cannot be negative and NEC condition (Strong Energy Condition). Finally, the energy density cannot be negative and NEC condition (Weak Energy Condition). In Figure \ref{Figure1}, we analyze these conditions graphically for the solutions previously found.

As observed in graph (a) the conditions are not violated, since  $\alpha=0$ and we have the classical electromagnetism case. In graph (b), when we include contributions from $f(R)$ gravity - with  $\alpha \neq 0$ - some energy conditions are violated. In graphs (c) and (d), some energy conditions are already violated, because we have regular geometries whose sources are nonlinear electrodynamics. By including $f(R)$ gravity corrections to these solutions in (d) and (f)  - with  $\alpha \neq 0$ - we notice that these energy conditions are still violated. Therefore, $f(R)$ gravity did not prevent the violation of energy conditions.

\section{A Regular Geometry}
Now we are interested in knowing whether, as in General Relativity, given a regular geometry, the energy conditions will be violated. Therefore, let us consider a hypothetical scenario in which we have a regular geometry within the context of $f(R)$ gravity. However, it turns out to be singular in the case of General Relativity:
\begin{equation}\label{2k(r)}
b(r) = -m + \frac{r^2}{l^2} -q^2\ln{\left(\alpha^2 l^2+\frac{r^2}{l^2}\right)}, 
\end{equation}
where $\alpha$ is the parameter related to the correction to $R$ in the $f(R)$ model.
The Ricci and Kretschmann scalars are given by
\begin{equation}
R = \frac{2 \left(-3 \alpha ^4 l^8+3 \alpha ^2 l^6 q^2-6 \alpha ^2 l^4 r^2+l^2 q^2 r^2-3 r^4\right)}{l^2 \left(\alpha ^2 l^4+r^2\right)^2}.    
\end{equation}
\begin{equation}
K = 4 \left[\frac{3}{l^4}+\frac{q^4 \left(3 \alpha ^4 l^8+2 \alpha ^2 l^4 r^2+3 r^4\right)}{\left(\alpha ^2 l^4+r^2\right)^4}-\frac{2 q^2 \left(3 \alpha ^2 l^4+r^2\right)}{l^2 \left(\alpha ^2 l^4+r^2\right)^2}\right].    
\end{equation}
These functions exhibit well-defined properties across their entire domain, ensuring a regular geometry, notably evident when approaching $r\to 0$. However, the regularity is compromised in the scenario where $\alpha=0$, namely, when we return to General Relativity. Following, we can perform an inversion to obtain $r(R)$ and integrate the equation (\ref{f(R)}) to determine the associated $f(R)$ gravity \cite{Rodrigues:2015ayd}:
\begin{eqnarray}
f(R) = R-2\Lambda + \frac{1}{4 \alpha l^8 q^2}[(\sqrt{l^4 q^4+4 \alpha ^2 l^6 q^2 \left(l^2 R+6\right)}-l^2 q^2)^2 \nonumber \\ \sqrt{\frac{(l^2 q^2 - l^4 (6 + l^2 R) \alpha^2 + \sqrt{l^4 q^4 + 4 l^6 q^2 (6 + l^2 R) \alpha^2)}}{(6 + l^2 R)}} -16 \alpha  l^6 q^4 \tan ^{-1}\beta(R)],
\end{eqnarray}
where,
\begin{equation}
\beta(R) = \frac{1}{\alpha  l^2} \sqrt{\frac{l^2 q^2-\left(\alpha ^2 l^4 \left(l^2 R+6\right)\right)+\sqrt{l^4 q^4+4 \alpha ^2 l^6 q^2 \left(l^2 R+6\right)}}{l^2 R+6}}
\end{equation}
and derivating with respect to $R$, we obtain
\begin{equation}
f_R(R) = 1 + \alpha  \sqrt{\frac{L^2 q^2-\left(\alpha ^2 L^4 \left(L^2 R+6\right)\right)+\sqrt{L^4 q^4+4 \alpha ^2 L^6 q^2 \left(L^2 R+6\right)}}{L^2 R+6}},  
\end{equation}
and as expected $\displaystyle f_R(R) = 1$, when $\alpha \to 0$. With the equation(\ref{condI}) and the $k(r)$ function of (\ref{2k(r)}), we can find the nonlinear electrodynamics associated
\begin{equation}\label{L(r)}
L(r) = \frac{q^2}{4 \pi } \left(\frac{\left(r^2-\alpha ^2 l^4\right)-2 \alpha ^3 l^4 r}{2 \left(\alpha ^2 l^4+r^2\right)^2}+\frac{\cot ^{-1}\left(\frac{\alpha  l^2}{r}\right)}{l^2}\right)+d_1,  
\end{equation}
where $d_1$ is a constant. Unfortunately, the equation (\ref{L(r)}) does not actually represent a nonlinear electrodynamics, since 
\begin{equation}
E(r) = \int \frac{r L'(r)}{4q} dr = \frac{q}{32 \pi } \left(\frac{2 c r^3-\alpha ^5 l^8-3 \alpha ^3 l^4 r^2}{\left(\alpha ^2 l^4+r^2\right)^2}+\alpha  \ln \left(\alpha ^2 l^4+r^2\right)\right),    
\end{equation}
and in the asymptotic limit, $r \to \infty$
\begin{equation}
E(r) \approx  \frac{\alpha  q \ln (r)}{16 \pi }.    
\end{equation}
Which is different from classical electromagnetism. This conclusion reinforces the idea that it seems in a $f(R)$ gravity, we do not have regular solutions sourced by nonlinear electrodynamics in $(2+1)$ dimensions. 

\subsection{Energy Conditions}
\begin{figure}[!h]
\begin{center}
\begin{tabular}{ccc}
\includegraphics[height=5cm]{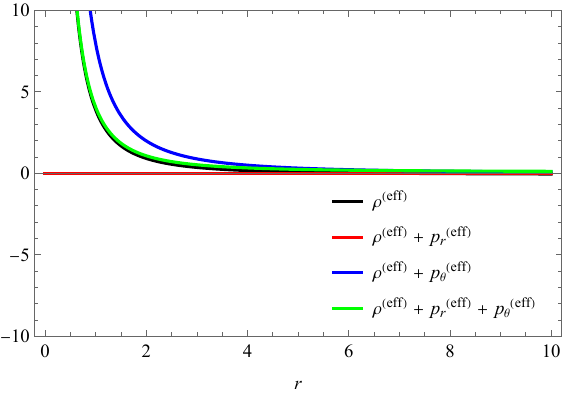}
\includegraphics[height=4.9cm]{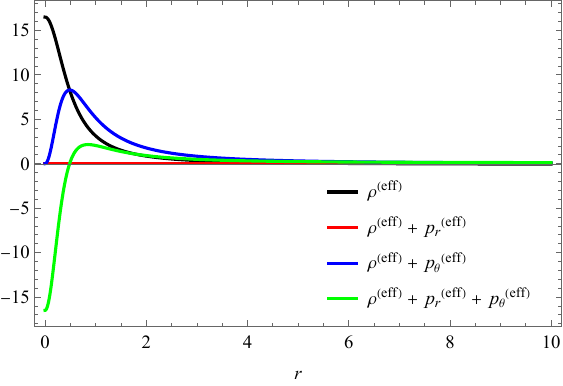}\\ 
(a) \hspace{8 cm}(b)
\end{tabular}
\end{center}
\caption{
The graphical representation of the radial dependence for $\rho^{(eff)}$ and combinations with $p_{r}^{(eff)}$ and $p_{\theta}^{(eff)}$ given from solutions (\ref{2k(r)}) with $l=0.7, q=10, m=80, \alpha = 0$ (left panel) and $l=0.7, q=10, m=80, a=2, \alpha = 1$ (right panel).\label{Figure2}}
\end{figure}

With the functions $f(R)$ and $L(r)$ defined, we graphically examine the energy conditions from the solution (\ref{2k(r)}) in Figure (\ref{Figure2}). In graph (a), the energy conditions are not violated, since for $\alpha \neq 0$ we return to the charged BTZ solution. However, when we regularize the solution with a term from $f(R)$ gravity, we observe that the energy conditions are violated. Thus, in principle, $f(R)$ gravity, even when exhibiting regularity, does not ensure the fulfillment of energy conditions.

\section{Conclusion \label{conclusion}}

In summary, we identified the necessary conditions to find a regular solution in $(2+1)$ dimensions supported by non-linear electrodynamics in any $f(R)$ gravity theory. We obtained a charged BTZ-like solution in $f(R)$ gravity and investigated two models of non-linear electrodynamics that provided regular solutions in General Relativity. We found that all solutions are non-regular. Therefore, the corrections in $f(R)$ gravity do not "cure" the singularity of the charged BTZ solution and it remove the regularity that non-linear electrodynamics had conferred in General Relativity. When analyzing the energy conditions, we observed that the energy conditions continue to be violated.

Attempting to answer the question: given a hypothetical geometry regularized by $f(R)$ gravity, would the energy conditions still be violated? We considered a regular geometry for $\alpha \neq 0$ and non-regular for $\alpha = 0$. We defined the $f(R)$ gravity, as well as the associated non-linear electrodynamics (since the electric field did not recover classical electromagnetism at infinity, regularity would not be effectively conferred by non-linear electrodynamics). We found that even in this scenario, the energy conditions would still be violated.

\section*{Acknowledgments}
\hspace{0.5cm} The authors thank the Funda\c{c}\~{a}o Cearense de Apoio ao Desenvolvimento Cient\'{i}fico e Tecnol\'{o}gico (FUNCAP), the Coordena\c{c}\~{a}o de Aperfei\c{c}oamento de Pessoal de N\'{i}vel Superior (CAPES), and the Conselho Nacional de Desenvolvimento Cient\'{i}fico e Tecnol\'{o}gico (CNPq), Grants no. 88887.822058/2023-00 (ACLS), no. 308268/2021-6 (CRM), and no. 200879/2022-7 (RVM) for financial support. R. V. Maluf acknowledges the Department de F\'{i}sica Te\`{o}rica de la Universitat de Val\`{e}ncia for the kind hospitality.



\end{document}